\documentclass[12pt,fleqn]{article}
\usepackage{epic,eepic}
\usepackage{graphicx}
\usepackage{epstopdf}

\usepackage{lineno}
\usepackage{gensymb}
\usepackage{color}

\begin{document}
\title{Thermodynamic metric geometry of the two-state ST2 model for supercooled water}

\author{Peter Mausbach,\footnote{Technical University of Cologne, Cologne, Germany (pmausb@gmx.net)} \, Helge-Otmar May,\footnote{University of Applied Sciences, Darmstadt, Germany, deceased} \, and George Ruppeiner\footnote{New College of Florida, Sarasota, Florida, USA}}

\maketitle
\begin{abstract}

Liquid water has anomalous liquid properties, such as its density maximum at 4\degree C. An attempt at theoretical explanation proposes a liquid-liquid phase transition line in the supercooled liquid state, with coexisting low-density (LDL) and high-density (HDL) liquid states. This line terminates at a critical point. It is assumed that the LDL state possesses mesoscopic tetrahedral structures that give it solid-like properties, while the HDL is a regular random liquid. But the short-lived nature of these solid-like structures make them difficult to detect directly. We take a thermodynamic approach instead, and calculate the thermodynamic Ricci curvature scalar $R$ in the metastable liquid regime. It is believed that solid-like structures signal their presence thermodynamically by a positive sign for $R$, with a negative sign typically present in less organized fluid states. Using thermodynamic data from ST2 computer simulations fit to a mean field (MF) two state equation of state, we find significant regimes of positive $R$ in the LDL state, supporting the proposal of solid-like structures in liquid water. In addition, we review the theory, compute critical exponents, demonstrate the large reach of the MF critical regime, and calculate the Widom line using $R$.

\end{abstract}


\section{Introduction}

Liquid water is unique among other pure fluids. Unlike ordinary fluids, cold water is known for its anomalous behavior, including a density maximum at 4\degree C and ambient pressure. Anomalous behavior is present as well in the isothermal compressibility $k_{T}$, the isobaric heat capacity $c_{p}$, and the thermal expansion coefficient $\alpha_{p}$ \cite{Debenedetti2003, Debenedetti2003a}. Both $k_{T}$ and $c_{p}$ increase significantly as water is cooled into the metastable liquid state, while $\alpha_{p}$ decreases. We focus here on the idea that the anomalous behavior of water finds its explanation in a second critical point (LLCP) in the metastable region, terminating a first-order liquid-liquid phase transition line (LLPT) between a high-density liquid state (HDL) and a low-density liquid state (LDL).

\par
Various scenarios have been proposed to explain the anomalous behavior of supercooled water \cite{Speedy1982, Poole1992, Sastry1996}. The possibility of a second critical point has attracted much attention \cite{Poole1992}. This scenario links water's anomalies to critical point fluctuations. Microscopically, the source of water's anomalies appear to be the presence of open tetrahedrally-coordinated networks of water molecules held together by hydrogen bonds (HB). It is believed that the competition between two polymorphic configurations of local molecular order generates the anomalies in the thermodynamic response functions of cold and supercooled water. In this picture, the HDL exhibits a high entropy local tetrahedrally coordinated HB network structure that is not fully developed. But the LDL develops low-entropy configurations consisting of open ``ice-like'' or ``solid-like'' HB network structure. At high temperatures and pressures, we would expect the HDL to dominate, leading to ordinary liquid behavior. But on cooling, the LDL could increasingly assert itself, with less conventional behavior.

\par
In two recent studies \cite{Ruppeiner2015, May2015}, the anomalous behavior of cold and supercooled liquid water was investigated by means of a relatively new approach, the thermodynamic metric geometry. This geometric concept has been systematically developed for atomic and molecular fluid systems using thermodynamic data obtained from experiments and computer simulations \cite{Ruppeiner2012, Ruppeiner2012a, May2012, May2013, Ruppeiner2016, Ruppeiner2017, Maus2018}.

\par
The resulting thermodynamic Ricci curvature scalar $R$ is of special interest. $R$ reveals information about the intrinsic physics of the considered system, and the size of organized mesoscopic structures \cite{Ruppeiner1979, Ruppeiner2010, Ruppeiner2013}. By applying thermodynamic fluctuation theory \cite{Ruppeiner1995}, $R$ can be formulated as a thermodynamic invariant, the same for a given thermodynamic state no matter what thermodynamic coordinates are used to calculate it. One of the main results of thermodynamic metric geometry is the hypothesis that the sign of $R$ specifies whether the intermolecular interactions are effectively attractive ($R < 0$) or repulsive ($R > 0$) (using the curvature sign convention of Weinberg \cite{Weinberg1972}). 

\par
The idea of the two-state nature of liquid water requires an unambiguous determination of both the presence and the proportion of each of the HDL and the LDL states, and this has come with partial success until recently \cite{Shiratani1996, Shiratani1998, Chau1998, Erring2001, Olein2006, Accord2011}. Because of the character of the thermodynamic Ricci curvature scalar $R$, investigations of significant regimes of either positive or negative $R$ allow for an identification of HDL or LDL states. This procedure was applied extensively in the studies of real water \cite{Ruppeiner2015, May2015}.

\par
Using the thermodynamic properties in the IAPWS-95 formulation \cite{Wagner2002} it was shown \cite{Ruppeiner2015} that stable cold water near the triple point displays a slab-like feature of positive $R$ in $(T,p)$ space, encompassing the density maximum at 4\degree C and ambient pressure. Experimental findings \cite{Malla2012, Malla2013, Malla2014} indicate that the $R > 0$ slab can be associated with an onset of HB clustering of open ice-like structures within the HB network \cite{Ruppeiner2015, Tanaka2012}. This connects to the concept of thermodynamic metric geometry since it was proposed \cite{Ruppeiner2015} that solid-like properties reveal themselves thermodynamically through positive values of the thermodynamic curvature $R$.

\par
We recently extended these geometric ideas into the metastable water regime \cite{May2015}. By applying a two-state equation of state (TSEOS), developed by Holten {\em et al.} \cite{Holten2014}, a dramatically decreasing $R$, to $-\infty$ at the postulated LLCP, was found on cooling into the metastable liquid state \cite{May2015}. On the LDL side of the LLPT line, organized ice-like structures led to positive $R$, whereas the HDL side shows only negative $R$. Positive $R$ in the LDL state appeared because tetragonal ice-like structures generally take up more space than the more disorganized HDL states. However, beyond the homogeneous ice nucleation limit the metastable liquid state is difficult to access experimentally due to the rapid homogeneous nucleation of ice. Therefore, the possibility of a LLPT in real water has to be examined from the extrapolation of properties far away from the LLPT.

\par
In principle, the formation of ice could be inhibited with computer simulations of water-like models, which allow deeper supercooling. Various potential models have been applied to the study of supercooled water, resulting in the discussion about the location of a possible LLCP. Table \ref{PotMod} provides a summary of some of these models. We adopt the term {\it apparent divergent point} (ADP), introduced by Pathak {\em et al.} \cite{Pathak2016}, in Table \ref{PotMod} to denote that several response functions appear to diverge at a specific point in $(T, p)$ coordinates. However, the existence of real criticality is uncertain. A true critical point reveals itself by a diverging correlation length and by specific values of the critical exponents, both very challenging to establish.

\par
Clear evidence of an LLPT has been reported \cite{PalmerNat2014} only for the ST2 model \cite{Still1974}, for which finite-size scaling \cite{Kesselring2013} confirms the existence of an LLCP. Only a weak divergence could be observed for other models such as the extended simple point charge (SPC/E) \cite{Berend1987} or the TIP4P potential \cite{Jorgen1983}, while the possibility of an ADP for the TIP5P potential \cite{Mahoney} is supported by recent simulation studies \cite{Uralcan2019}. For the mW model \cite{Molin2009} it has been shown \cite{Moore2011} that the non-ideality in mixing two alternative local orders is entropy-driven, and that this is too weak to produce an LLPT. The situation for the TIP4P/2005 potential \cite{Abascal2005} is unclear, with evidence for an LLPT reported \cite{Abascal2010, Biddle2017}, but also questioned in other work \cite{Overduin2013}. An LLCP was also predicted by using an energy landscape analysis \cite{Handle2018}. The E3B3 water potential \cite{Taint2015} explicitly includes three-body interactions superimposed on the TIP4P/2005 model as its two-body reference. Ni and Skinner \cite{Ni2016} present some evidence that this model shows an ADP. The iAMOEBA potential \cite{Wang2013} is a classical polarizable water model. It is highly accurate for modelling water in the solid and fluid phases. Pathak {\em et al.} \cite{Pathak2016} report that an ADP exists for this potential. A deeper discussion of the model's strengths and weaknesses may be found in refs. \cite{Pathak2016, Gallo2016, Palmer2018}.

\begin{table}
\begin{center}
\begin{tabular}{l||c|c|c|c|c}
potential & $T_{\rm \:ADP}$ (K) & $p_{\rm \:ADP}$ (MPa) & Ref. \\[0.5ex] \hline
SPC/E & 130 & 290 & \cite{Pathak2016} \\
iAMOEBA & 175 & 184 & \cite{Pathak2016} \\
ST2 & 246 & 206 & \cite{Kesselring2013} \\
TIP5P & 213 & 340 & \cite{Uralcan2019} \\
mW  & - & - & \cite{Moore2011} \\
TIP4P/2005 & 175 & 175 & \cite{Handle2018} \\
E3B3  & 180 & 210 & \cite{Ni2016} \\

TIP4P & 190 & 150 & \cite{Corrad2010} \\
\end{tabular}
\end{center}
\caption{\label{PotMod} Potential models applied to the LLPT hypothesis of water. The table contains the ($T, p$) coordinates of the ADP (apparent divergent point, as described in the text) for the potentials, as well as references containing data.} 
\end{table}

\par
In a recent contribution by Anisimov {\em et al.} \cite{Anisim2018}, a generic phenomenological approach was introduced describing water and water-like phenomena without the need for postulating a potential model. In this generic approach, the Gibbs energy is formulated for a TSEOS in which the ``background" term (cf. chapt. 2) is replaced by a lattice-gas. This approach is intended to unify the debated scenarios.

\par
However, in the present study we are interested in the geometric behavior of concrete water models for which, presently, a high quality EOS, e.g. of the Holten-Anisimov-Sengers type \cite{Holten2014}, is necessary. Based on the two-structure concept, TSEOS's were developed so far only for the mW \cite{Holten2013}, ST2 \cite{Holten2014a} and the TIP4P/2005 \cite{Biddle2017} potential. The TSEOS of Biddle {\em et al.} \cite{Biddle2017} was applied for exploring the doubly metastable region, where liquid water is both supercooled and under tension. With our focus on systems with a clear evidence for an LLCP we chose the ST2-TSEOS \cite{Holten2014a} as a starting point for investigations, accepting that the potential exhibits a number of quantitative deviations from the behavior of real water \cite{Poole2005}. The use of the ST2 potential has a long tradition \cite{Geiger1979, Blum1984, Geiger1991} and the proposal of an LLCP in supercooled water emerged from computer simulation studies of this model \cite{Poole1992}.

\par
The ST2 model comes in two versions, ST2(I) and ST2(II). Based on simulation studies \cite{Liu2012, Poole2013} both versions have been fit to a mean-field (MF) thermodynamic picture, and MF has been augmented by a crossover procedure to critical behavior \cite{Holten2014a}. In this paper we compute $R$ for the MF pictures, ST2(I-MF), and ST2(II-MF), both of which have non-zero critical pressures, as opposed to the model we employed in our previous study \cite{May2015} that has its second critical point at zero pressure.

\par
The paper is organized as follows. First, the two-state thermodynamics and the character of the associated EOS are summarized. Then, we explain how we calculate $R$ from the TSEOS. An analysis and discussion of results for $R$ in the supercooled water region follows. This discussion includes power law analyses of thermodynamic quantities near the critical point. We also key on the location of the Widom line corresponding to curves of maximum $|R|$ at constant pressure.

\section{Two-state thermodynamics of water}

The anomalous behavior of several thermodynamic properties of metastable supercooled water has motivated various attempts at theoretical explanation in terms of so-called two-state models that have been applied to thermodynamic phenomena in one-component liquids \cite{Rapoport1967, Aptekar1968, Ponyatovsky1998, Tanaka2000, Tanaka2000a, Angell2000, Cuthbertson2011}. It is assumed that liquid water at low temperatures can be described as a mixture of two states, a HDL state (index A) and a LDL state (index B). The fraction of water molecules in state B is denoted by $x$ ($x\in[0,1]$), and is controlled by a ``reaction'' $A\rightleftharpoons B$.

\par
In our present study, the geometric structure of the ST2 model for supercooled water is analyzed based on the TSEOS developed by Holten {\em et al.} \cite{Holten2014a}, which models well the properties of real water \cite{May2015, Holten2014}. Two versions of the TSEOS have been developed \cite{Holten2014a}, a mean-field (MF) description and a so-called crossover (CO) approach, accounting for critical order-parameter fluctuations.

\par
For the MF approach we start with the molar Gibbs free energy $G=G(T,p,x)$, where $T$ and $p$ denote the temperature and pressure, respectively. We adopt the following expression for the two-state mixture \cite{Holten2014a}:

\begin{equation}
G = G^A + x G^{BA} + R_G T \left[x \ln x + (1 - x)\ln(1 - x) + W x(1 - x)\right]\label{10},
\end{equation}

\noindent where $G^A$ is the molar Gibbs free energy of the pure state $A$, $G^{BA}=G^B-G^A$, with $G^B$ the molar Gibbs free energy of the pure state $B$, $W$ is the measure of the nonideality of mixing, and $R_G$ is the universal gas constant. $G^A$, $G^{BA}$, and $W$ are functions of $T$ and $p$ but not of $x$. $G^{BA}$ is related to the equilibrium constant $K$ of the reaction $A\rightleftharpoons B$:

\begin{equation}
\ln K = - \frac{G^{BA}}{R_G T}.
\label{20}
\end{equation}

\noindent Along the LLPT, $G^{BA}=0$. Thus, the condition $\ln K=0$ determines the LLPT.

\par
The molar fraction $x$ is unconstrained by any conservation law, so in this model we take $(T, p)$ to be given, and let $x$ float so as to minimize $G$:

\begin{equation}\left(\frac{\partial G}{\partial x}\right)_{T, p}=0.\label{30}\end{equation}

\noindent This yields the equilibrium value $x=x_e$, where $x_e$ is the numerical solution to

\begin{equation}
\ln K - \ln\left(\frac{x}{1-x}\right) - W\times(1-2 x) = 0.
\label{40}
\end{equation}

\par
In the sequel, we use the following dimensionless quantities:

\begin{equation}
\hat{G} = \frac{G}{R_G T_C}, \; \;
\hat{T} = \frac{T}{T_C}, \; \; \;
\tau = \frac{T-T_C}{T_C}, \; \; \;
\pi = \frac{p-p_C}{R_G T_C\,\rho_C},
\label{50}
\end{equation}

\noindent where $T_C$, $p_C$, and $\rho_C$ denote the critical temperature, pressure, and density, respectively. The dimensionless molar Gibbs free energy $\hat{G}^A=G^A/R_G T_C$ of the pure component $A$ defines the ``background" that must be explicitly entered into the model. For example, we use

\begin{equation}
\hat{G}^A = \sum_{m, n} c_{mn}\ \tau^{m} \pi^{n},
 \label{60}
\end{equation}

\noindent where $m$ and $n$ are integers, and the $c_{mn}$ are suitable constants, with only $c_{00}, c_{01}, c_{02}, c_{03}, c_{11}, c_{12}, c_{13}, c_{20},$ and $c_{30}$ nonzero in \cite{Holten2014a}.

\par
Physically valid roots of Eq. (\ref{40}) for $x=x_e(T, p)$ must be in the range $x_e\in [0,1]$. For the great majority of the $(T,p)$ points in our evaluation grids, there was one and only one root in this range, and it always corresponded to a local minimum of $G$ with respect to variations in $x$. For the remainder of the points, all with $T<T_C$, there were three roots, with the middle one corresponding to a local maximum of $G$. The other two roots corresponded to local minima of $G$, and are indicators of the two possible phases. The correct physical root is the one with the smaller $G$.

\par
Holten {\em et al.} \cite{Holten2014a}, assumed that $\ln{K}$ follows the linear expression

\begin{equation}
\ln K =\lambda(\tau+a\,\pi),
\label{70}
\end{equation}

\noindent where $\lambda$ and $a$ are fit parameters. The LLPT, and its analytic continuation the Widom line,\footnote{The Widom line was defined as the ``locus of maximum correlation length'' by Franzese and Stanley \cite{Franzese2007}, a definition much seen in the literature. Earlier, however, Griffiths and Wheeler \cite {Griffiths1970} referred to the concept of the Widom line as the ``linear extension of the coexistence curve in the $p-T$ plane.'' This concept may be challenged since the behavior at the critical point is nonanalytic beyond MF theory. Widom and Rowlinson \cite{Widom1970} focussed on the critical isochor and refer to ``either the critical isochor or the locus on which $\left(\partial^2 p/\partial\rho^2 \right)_{T}=0$ above $T_C$.'' Holten {\em et al.} \cite{Holten2014a}, in their MF context, take the Widom line to be the analytic continuation of the phase transition line, and this is the picture that we feature here.} are given by

\begin{equation} \tau+a\,\pi = 0. \label{75}\end{equation}

\noindent It has been shown \cite{Holten2014a} that the LLPT in ST2 is energy-driven, resulting in

\begin{equation}
W = \frac{2 + \omega_{r}\,\pi}{\hat{T}},
\label{80}
\end{equation}

\noindent where $\omega_{r}$ is an adjustable coefficient.

\par
Holten {\em et al.} \cite{Holten2014a} considered two versions of the ST2 model of water. The ST2(I) version employed the reaction field method to approximate electrostatic interactions \cite{Poole2005, Cuthbertson2011}. The ST2(II) version employed the Ewald treatment of electrostatics with vacuum boundary conditions \cite{Liu2012, Liu2009, Palmer2013}. For each version, both the MF and the CO approaches were applied, yielding four different models ST2(I-MF), ST2(I-CO), ST2(II-MF), and ST2(II-CO). The CO approach accounts for critical order-parameter fluctuations in the vicinity of the LLCP. We made a preliminary calculation, and found that the CO approach changed the results for $R$ only little outside the critical region. We will report details in a future publication, and focus in this paper just on the MF models: ST2(I-MF) and ST2(II-MF). Necessary fitting parameters are found in \cite{Holten2014a}.

\section{Thermodynamic metric geometry of two-state thermodynamics}

\par
The basis for the calculation of the Ricci thermodynamic curvature scalar $R$ is a line element $d\ell$ introduced by the thermodynamic entropy information metric \cite{Ruppeiner1995}

\begin{equation}
d\ell^2 = \sum_{i,j} g_{ij} dq^i dq^j.
\label{90}
\end{equation}

\noindent $d\ell^2$ is an invariant in the thermodynamic parameters $q^i$, and the coefficients $g_{ij}$ are the components of the thermodynamic metric tensor. For a one-component fluid there are two independent state variables $q^1$ and $q^2$ and the Ricci curvature scalar is calculated from \cite{Ruppeiner2012a, Sokolnikoff}

\begin{equation} \begin{array}{lr} {\displaystyle R= -\frac{1}{\sqrt{g}} \left[ \frac{\partial}{\partial q^1}\left(\frac{g_{12}}{g_{11}\sqrt{g}}\frac{\partial g_{11}}{\partial q^2}-\frac{1}{\sqrt{g}}\frac{\partial g_{22}}{\partial q^1}\right) \right. } \\ \hspace{3.6cm} + {\displaystyle \left. \frac{\partial}{\partial q^2}\left(\frac{2}{\sqrt{g}} \frac{\partial g_{12}}{\partial q^1} -\frac{1}{\sqrt{g}}\frac{\partial g_{11}}{\partial q^2}-\frac{g_{12}}{g_{11}\sqrt{g}}\frac{\partial g_{11}}{\partial q^1}\right)\right],} \end{array} \label{curvature}\end{equation}

\noindent with

\begin{equation}
g = g_{11}\, g_{22} - g_{12}^2.
\label{100}
\end{equation}

\noindent The value of $R$ for any thermodynamic state is independent of the coordinate system used to calculate it.

\par
Here, the independent state variables are $(q^1, q^2) = (T, p)$, and the thermodynamic metric elements become, in terms of the molar Gibbs free energy $G$, \cite{Ruppeiner2012a}

\begin{equation}
g_{11} = -\frac{1}{k_B T v} \frac {\partial^2 G}{\partial T^2},
\label{110}
\end{equation}

\begin{equation}
g_{22} = -\frac{1}{k_B T v} {{\frac {\partial ^{2} G}{\partial {p}^{2}}}},
\label{120}
\end{equation}

\noindent and

\begin{equation}
g_{12} = -\frac{1}{k_B T v} {{\frac {\partial ^{2} G}{\partial {p}\, \partial T}} },
\label{130}
\end{equation}

\noindent where

\begin{equation}v=\left(\frac{\partial G}{\partial p}\right)_{T}\label{133}\end{equation}

\noindent is the molar volume, and $k_B$ is Boltzmann's constant; $k_B=R_G/N_A$, with $N_A$ Avogadro's number.

\par
The isothermal compressibility is

\begin{equation}k_{T}=-\frac{1}{v}\left(\frac{\partial v}{\partial p}\right)_T.\label{133}\end{equation}

\noindent The isobaric molar heat capacity is

\begin{equation}c_{p}=T\left(\frac{\partial s}{\partial T}\right)_p.\label{135}\end{equation}

\noindent where

\begin{equation}s=-\left(\frac{\partial G}{\partial T}\right)_p\label{137}\end{equation}

\noindent is the molar entropy.

\par
The calculation of the derivatives of $G$ with respect to $(T,p)$ is complicated by the dependence of $x_e$ on $T$ and $p$. We cannot obtain an analytic expression for the equilibrium $x_e(T, p)$, since Eq. (\ref{40}) does not solve in closed form for $x_e$. To calculate $R$, we picked specific values of $(T,p)$, and numerically solved Eq. (\ref{40}) for $x_e(T,p)$. Numbers for the derivatives of $x_e(T,p)$ up to third-order result from the implicit differentiation of Eq. (\ref{40}). These numbers get substituted into the explicit expression for $R$ in terms of $x_e(T,p)$ and its derivatives. This procedure is thus very different in style from the calculation of $R$ based on the Wagner-Pru{\ss} equation \cite{Ruppeiner2015, Wagner2002} that used an explicit expression for the Helmholtz free energy.

\par
$R$ has some interesting physical properties. First, calculations in the critical region for: 1) pure fluids \cite{Ruppeiner1979}, 2) the one-dimensional ferromagnetic Ising model \cite{Ruppeiner1981}, 3) the one-dimensional Takahashi gas \cite{Ruppeiner1990}, 4) a decorated Ising chain \cite{RuppeinerBellucci}, and 5) a variety of other spin models \cite{Johnston2003}, all yielded a relationship between the curvature and the correlation length $\xi$:

\begin{equation}
\xi^d=\frac{|R|}{2},
\label{140}
\end{equation}

\noindent where $d$ is the spatial dimensionality. Second, it was found by calculation in a number of cases that the sign of $R$ indicates the nature of the microscopic interactions, with $R>0$ corresponding to effectively repulsive interactions, and $R<0$ to effectively attractive interactions \cite{Ruppeiner2010}. For a scenario where the interpretation of the sign of $R$ is not as clear cut; see Bra{\'n}ka {\it et al.} \cite{Branka2018}.

\par
The metric geometry of thermodynamics has also been applied in the black hole scenario \cite{Ruppeiner2013, Ruppeiner2014}.

\section{Results and Discussion}

The two-state ST2 model of Holten {\em et al.} was fit to simulation data. The range of the simulations is given in Figure 1 of \cite{Holten2014a}, and thus we plot the temperature from $240-340$ K, and the pressure from $100-250$ MPa. For ST2(I-MF), the critical point parameters are $T_C=253.5$ K and $p_C=160.0$ MPa, and for ST2(II-MF), $T_C=249.0$ K and $p_C=146.0$ MPa.

\subsection{Overall picture}

\par
\begin{figure}
\begin{minipage}[b]{0.5\linewidth}
\includegraphics[width=2.7in]{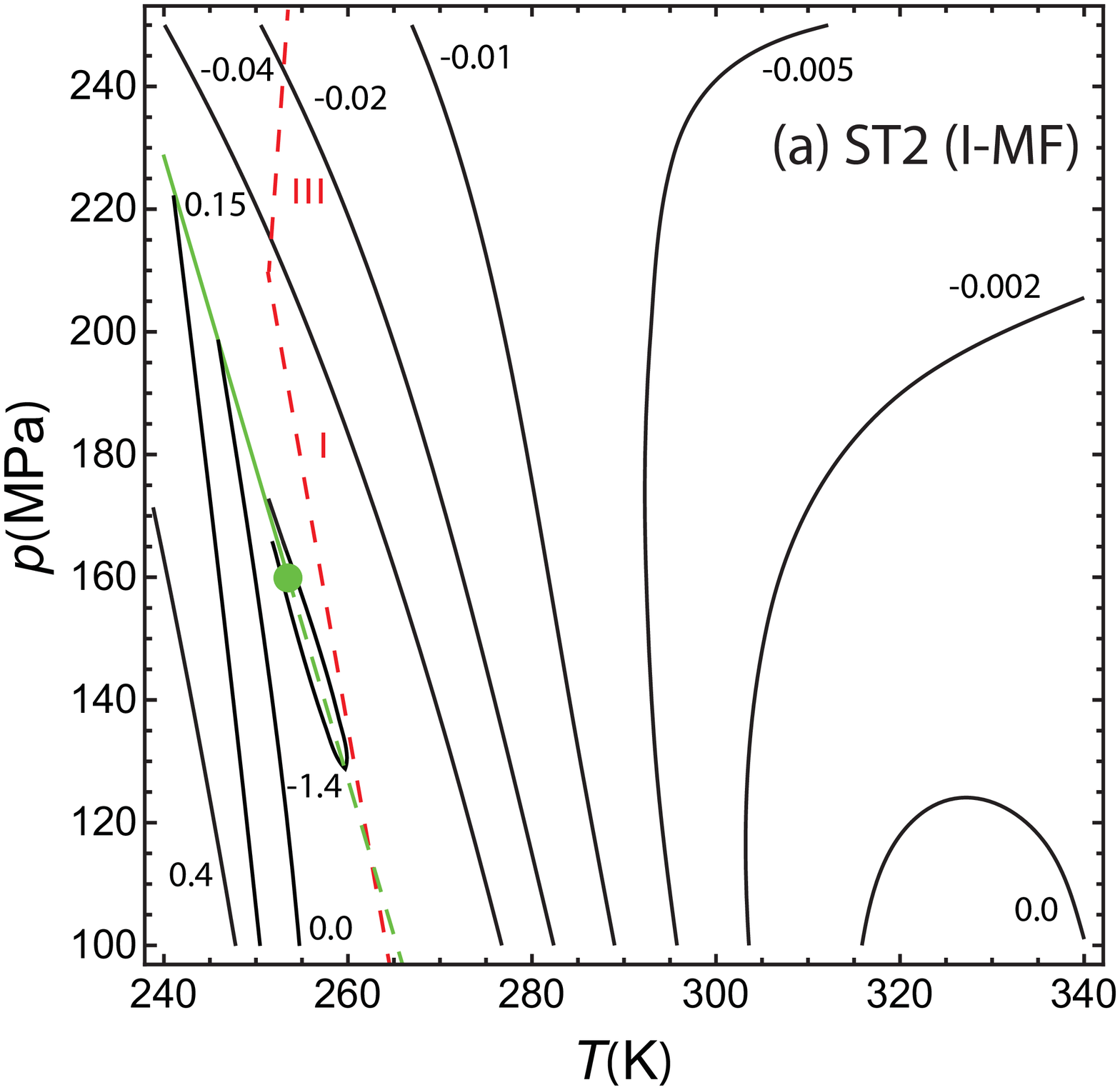}
\end{minipage}
\hspace{0.0 cm}
\begin{minipage}[b]{0.5\linewidth}
\includegraphics[width=2.7in]{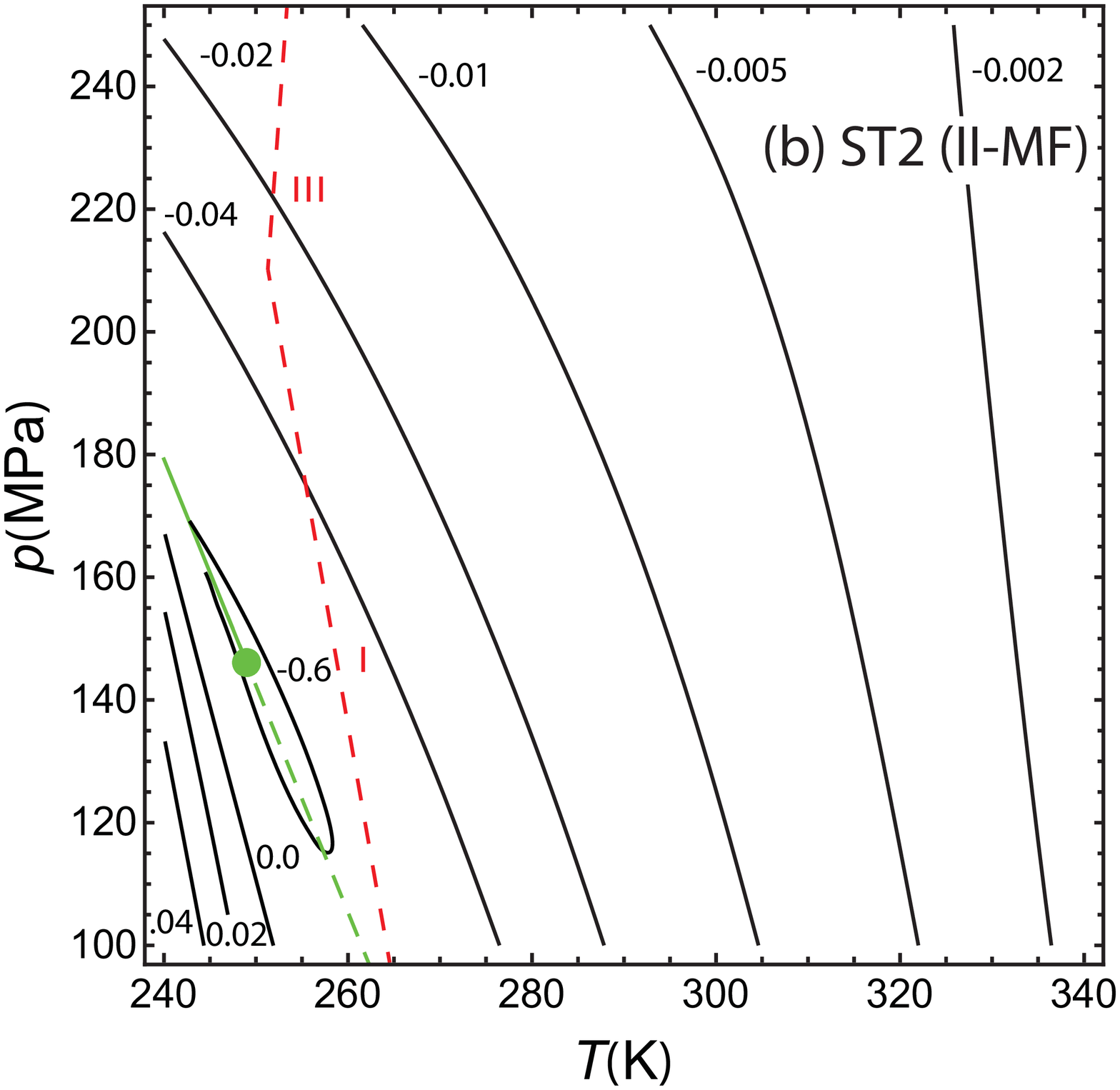}
\end{minipage}
\begin{center}
\includegraphics[width=2.7in]{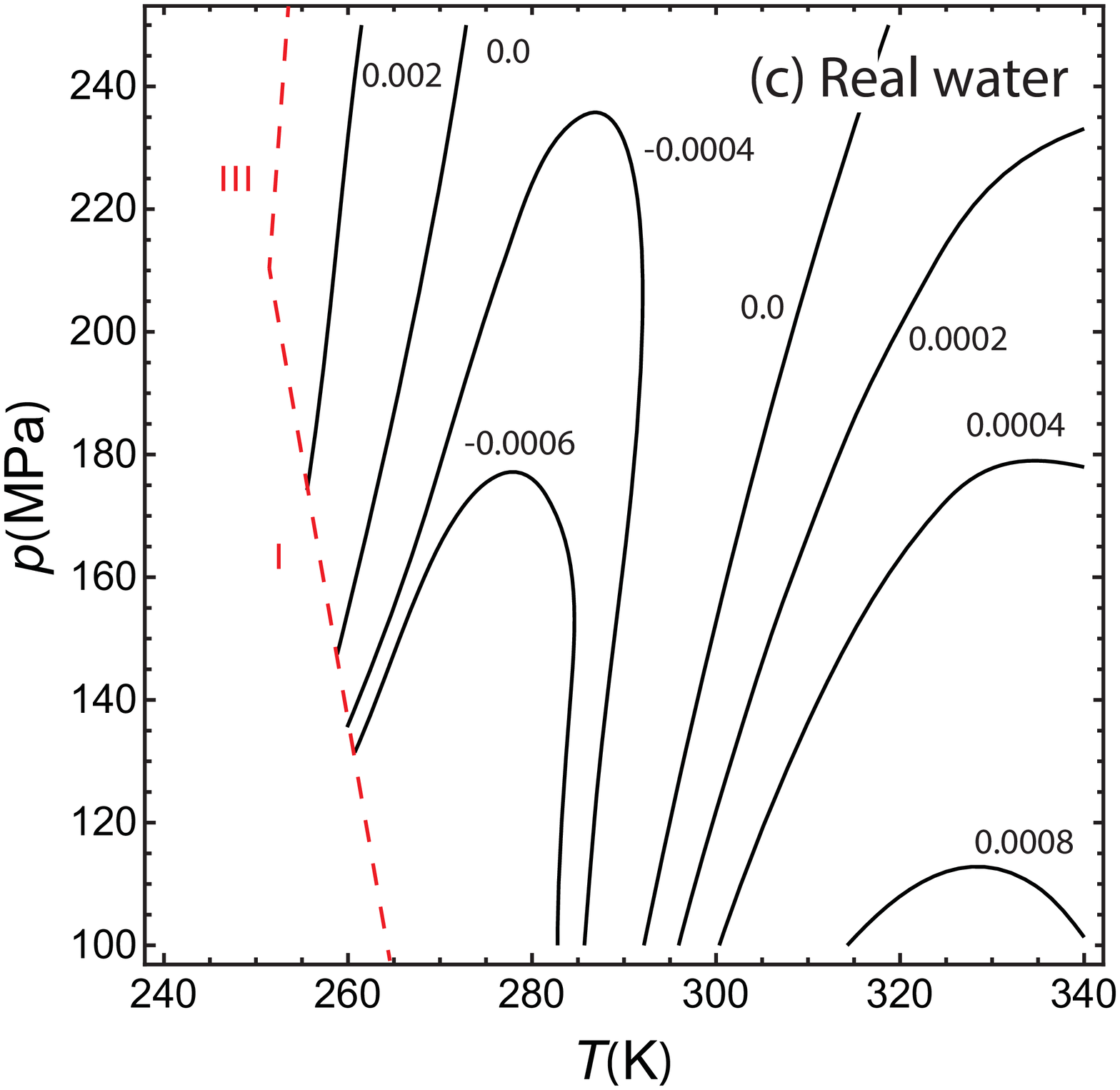}
\end{center}
\caption{$R$-diagrams ($R$-contours reported in nm$^3$ units) for (a) ST2(I-MF), (b) ST2(II-MF), and (c) stable water. The LLPT's are indicated by solid green curves terminating in circles at their LLCP's. The Widom lines are indicated by dashed green lines. The solid-liquid phase transition line in the stable phase is indicated by the dashed red line in each graph, with phases Ice I and III joined discontinuously. Both diagrams (a) and (b) show regions of positive $R$ in the LDL to the left of their LLPT lines. In addition, ST2(I-MF) mimics qualitatively the boundary of the positive $R$ region in the normal liquid encompassing the triple point. }
\label{figure1}
\end{figure}

In Figure \ref{figure1}, we show the $R$-diagrams ($R$-contours) for the ST2(I-MF) and ST2(II-MF) models. For comparison, we also show the corresponding $R$-diagram for real stable water \cite{Ruppeiner2015}. The units of $R$ are cubic nanometers per molecule (to set the scale, the cube of twice the Bohr radius is about $0.0012$ nm$^3$). The majority of the states in these figures represent normal water. However, using basic computer models to understand large-scale water properties is very challenging, and it gets harder the higher the order of the derivatives of $G(T,p)$. So, our comparison of the simulation $R$'s with those in real stable water is intended only to set a general context.

\par
Cooling the system isobarically into the supercooled state at pressure $p>p_C$ results in negative and decreasing $R$ as the LLPT is approached. The LLCP is looped by curves of constant negative $R$; for example, see the $R=-1.4$ nm$^3$ loop for ST2 (I-MF), and the $R=-0.6$ nm$^3$ loop for ST2(II-MF). $R$ gets more negative the closer we get to the LLCP, at which $R$ diverges to $-\infty$. Such behavior is consistent with all of the fluid critical point models known so far \cite{Ruppeiner2015, Ruppeiner2017}.

\par
$R$ has significant regimes of positive values for temperatures below the LLPT, indicating more ordered ice-like structures in significant portions of the LDL state. These positive $R$ values are considerably more pronounced for ST2(I-MF), with larger $R$ indicating more significant ice-like structures. In both cases, positive and negative values of $R$ are separated by $R=0$ lines extending out from the LLPT's in the directions of the critical points. Such extensions are also features of regular critical points in fluid systems with complex molecules; see Ruppeiner {\it et al.}, Figure 11(b) \cite{Ruppeiner2017}.

\par
For regular critical points, the low-density (vapor) phase may have positive or negative $R$'s depending on the presence or absence of some type of organization within the fluid. This certainly matches our thinking here about ice-like structures in the LDL state of supercooled water. For ST2(I-MF) in the lower right-hand corner of Fig. \ref{figure1}(a), the $R=0$ contour may indicate roughly the top of the positive slab of $R$ that was found independently in real water near ambient conditions \cite{Ruppeiner2015}. A portion of this slab is shown in Fig. \ref{figure1}(c). ST2(II-MF) in Fig. \ref{figure1}(b) does not show such a feature. Holten {\em et al.} \cite{Holten2014a} stated that their approximations, in particular Eq. (\ref{70}) here, make their EOS less accurate away from the LLCP. Nevertheless, qualitative similarities might be physically meaningful.

\par
In addition, the earlier water study shown in Fig. \ref{figure1}(c) found that $R$ becomes positive on approaching the Ice III phase, a phase spanning roughly a line from

$$(T, p):(256.2 \mbox{ K}, 209.9 \mbox{ MPa})\to (273.3 \mbox{ K}, 350.1 \mbox{ MPa}).$$

\noindent A portion of this Ice III phase transition line is shown in Fig. \ref{figure1}(c), with an $R=0$ line in the fluid fronting the Ice III curve. This intrusion of the effects of the solid phase on the negative $R$ contribution of the HDL may mark a limitation of the ST2 TSEOS.

\subsection{Phase transition line}

\par
\begin{figure}
\begin{minipage}[b]{0.5\linewidth}
\includegraphics[width=2.75in]{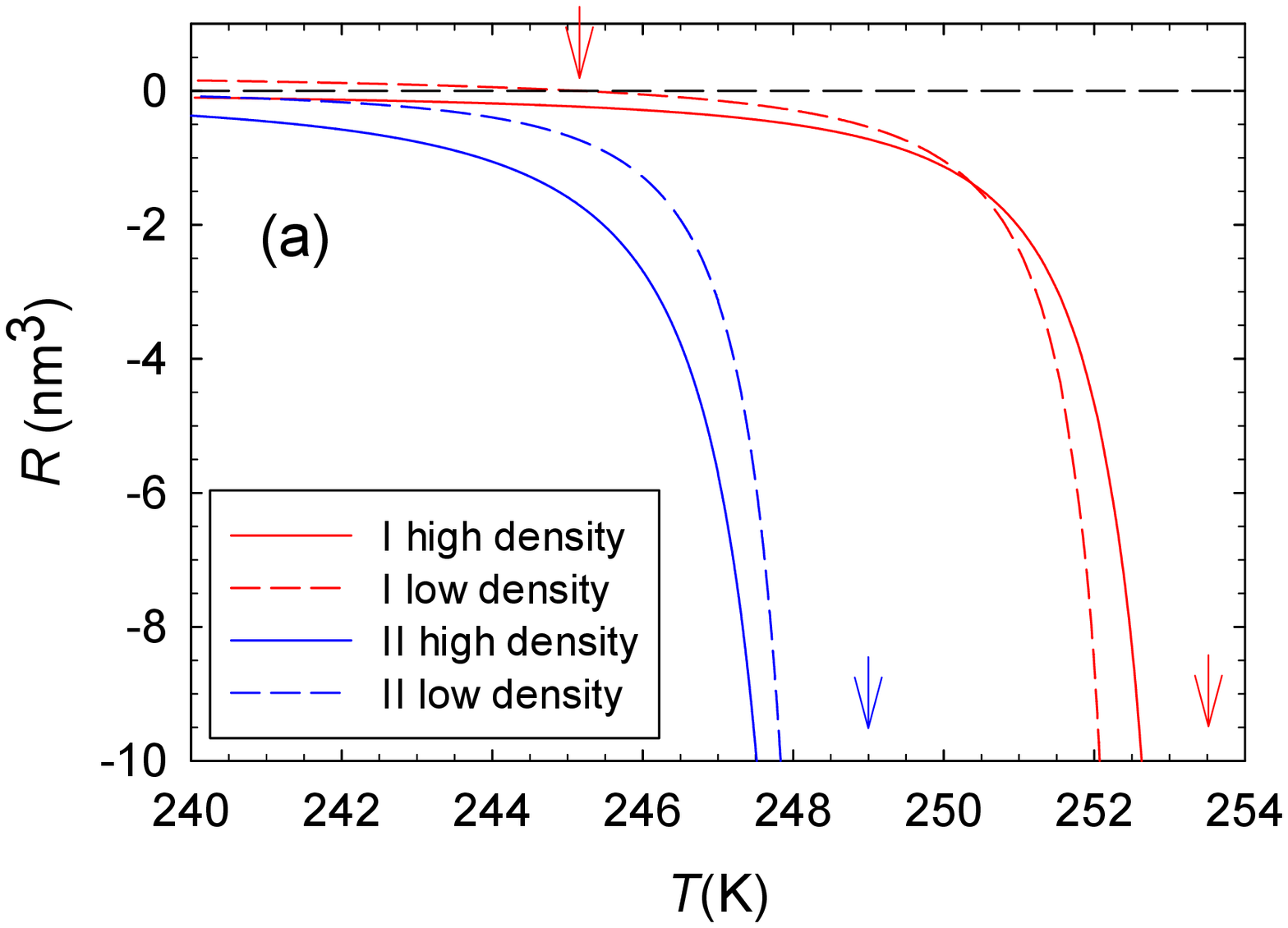}
\end{minipage}
\hspace{0.0 cm}
\begin{minipage}[b]{0.5\linewidth}
\includegraphics[width=2.81in]{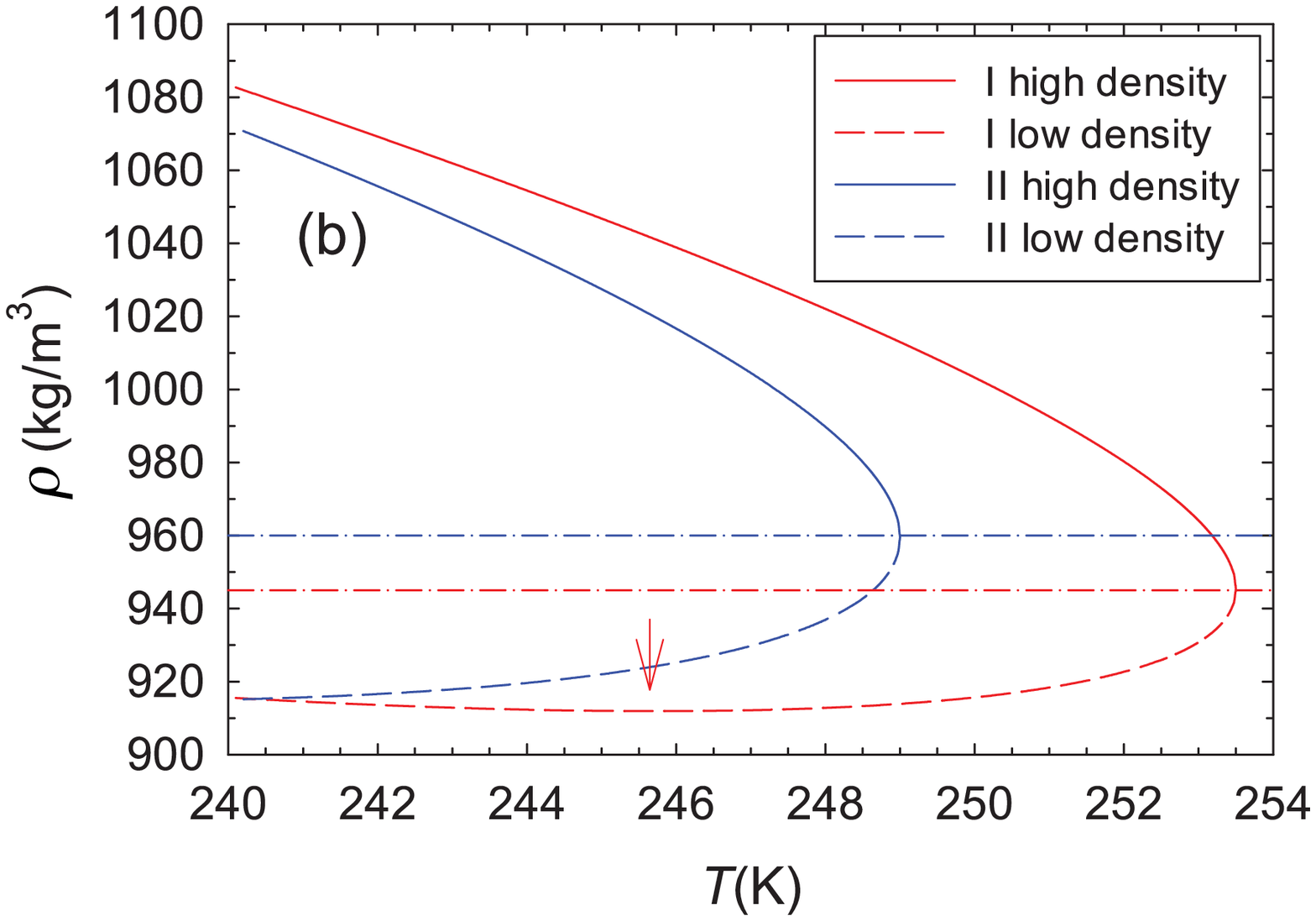}
\end{minipage}
\caption{Thermodynamic properties along the high and the low-density branches of the LLPT's for the ST2(I-MF) and ST2(II-MF) models. We show: (a) $R$ with negative and diverging critical point values, and with uniformly negative values in the high-density branches (with the $T_C$'s indicated by down arrows). For the ST2(I-MF) low-density branch, $R$ crosses to positive values for $T<245.21$ K. (b) $\rho$ receding from the respective critical density values (indicated by the straight dashed lines) on lowering $T$ from $T_C$. The ST2(I-MF) low-density branch has a density minimum at $T=245.73$, corresponding closely to the temperature of the sign change in $R$.}
\label{figure2} 
\end{figure}

Figure \ref{figure2}(a) shows the thermodynamic curvature $R$ versus the temperature $T$ along the LLPT's, given by Eq. (\ref{75}) in the branch with $\tau<0$. Near the critical points we see two features that appear to be general in pure fluids \cite{Ruppeiner2012a, Ruppeiner2017}: (1) all the four LLPT branches have negative $R$'s, diverging at their respective critical points, and (2) the two high-density branches both show negative values of $R$ over their full ranges. For ST2(I-MF), $R$ for the low-density branch crosses over to positive values on cooling below $T=245.21$ K. Generally, in the low-density branch of pure fluids composed of molecules having structures not too simple, such a change in sign is ubiquitous \cite{Ruppeiner2017}, so it is not surprising to see it here. For ST2(II-MF), this change in the sign of $R$ did not appear in the temperature regimes explored here, but it is expected at lower temperatures.

\par
Figure \ref{figure2}(b) shows the LLPT's in $(\rho, T)$ space, where $\rho=1/v$ is the density. As $T$ is lowered from $T_C$, $\rho$ recedes monotonically from the respective critical point values, except for the ST2(I-MF) low-density branch where $\rho$ reaches a density minimum at $T=245.73$ K. This density minimum coincides closely with the sign change in $R$. Holten {\it et al.} show this density minimum in their Figure 5 \cite{Holten2014a}. Earlier, Poole {\it et al.} also found a density minimum in ST2 simulations \cite{Poole2005}. The ST2(II-MF) model shows neither an $R$ sign crossing nor a density minimum in its low-density branch. But that may be because the temperature has not gone low enough. We explicitly verified that all of the states in Fig. \ref{figure2} are thermodynamically stable.

\subsection{Asymptotic behavior close to the LLCP}

In this subsection we discuss the asymptotic behavior of the ST2(I-MF) and ST2(II-MF) models near their critical points. MF critical exponent values are expected. Our discussion is numerical in nature, with little resort to rigorous critical phenomena theory.

\par
Questions posed in this subsection are: What are the calculated values of the critical exponents of the ST2(I-MF) and ST2(II-MF) models? How far out does the asymptotic MF critical point regime extend? What is the nature of the Widom line, and can we compute it using $R$, $k_{T}$, and $c_{p}$?

\subsubsection{Critical exponents}

A frequent measure of the distance to the critical point is the reduced temperature $\tau$ in Eq. (\ref{50}). In terms of $\tau$, we may write for some thermodynamic quantity $X$ the asymptotic power law expression

\begin{equation}X=X_0\left|\frac{T-T_C}{T_C}\right|^{-x},\label{}\end{equation}

\noindent where $X_0$ and $x$ are the critical amplitude and the critical exponent of $X$, respectively. The ``natural'' paths for approaching the critical point are the phase transition line, and it's logical continuation into the supercritical regime via the critical isochor, or the Widom line. For these paths, the critical exponents for the isochoric heat capacity and the isothermal compressibility are $\alpha=\alpha'$ and $\gamma=\gamma'$, respectively \cite{Pathria2011, Stanley1999}. Prime/no prime on the critical exponents denote $\tau$ negative/positive.

\par
A number of analyses have been carried out approaching the conjectured supercooled water critical point along isobars, an idea pioneered by Speedy and Angell \cite{Speedy1976}. The critical exponents along isobars will be different from those along the ``natural'' paths.

\par
We also have a critical exponent $\beta$ for the difference between the coexisting densities $\rho_{HDL}$ and $\rho_{LDL}$:

\begin{equation}\left(\frac{\rho_{HDL}-\rho_{LDL}}{\rho_C}\right)=\rho_0\left|\tau\right|^{\beta},\label{}\end{equation}

\noindent where $\rho_0$ is the critical amplitude. In addition, there is a critical exponent $\nu$ for the correlation length $\xi$. By the proportionality Eq. (\ref{140}), the critical exponent for $R$ is the product $d\,\nu\,$. The critical exponent for $R$ has been explicitly shown \cite{Ruppeiner2012a, Ruppeiner1979, Johnston2003} to obey the hyperscaling relation

\begin{equation}d\,\nu=2-\alpha\label{858373}.\end{equation}

\noindent For MF, the critical exponents are $\alpha=0$, $\beta=1/2$, and $\gamma=1$, all values independent of $d$ \cite{Pathria2011}. With $\alpha=0$, the critical exponent for $R$ is $2$, independent of $d$.

\par
\begin{figure}
\begin{minipage}[b]{0.5\linewidth}
\includegraphics[width=2.8in]{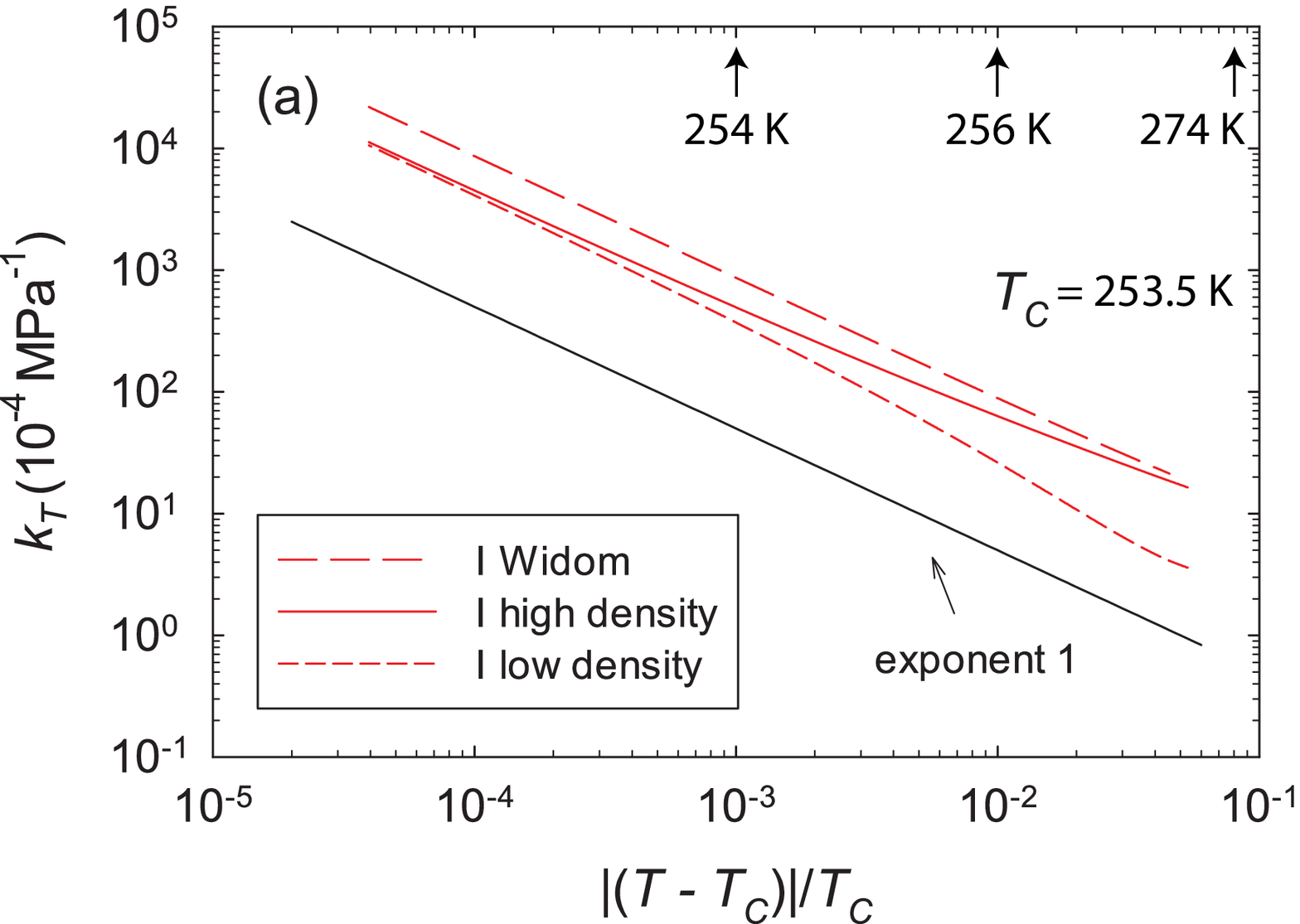}
\end{minipage}
\hspace{0.0 cm}
\begin{minipage}[b]{0.6\linewidth}
\includegraphics[width=2.8in]{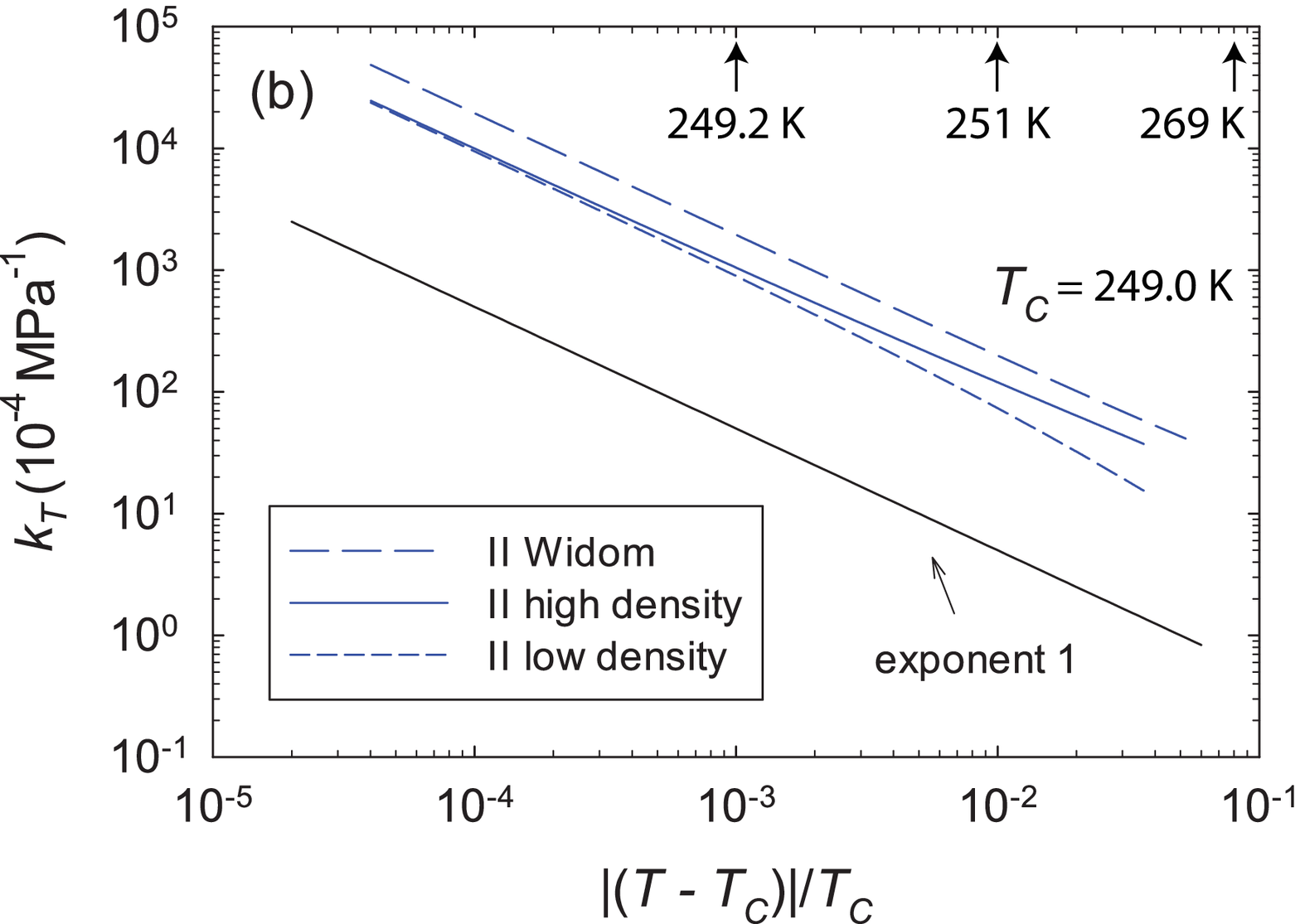}
\end{minipage}
\begin{minipage}[b]{0.5\linewidth}
\includegraphics[width=2.8in]{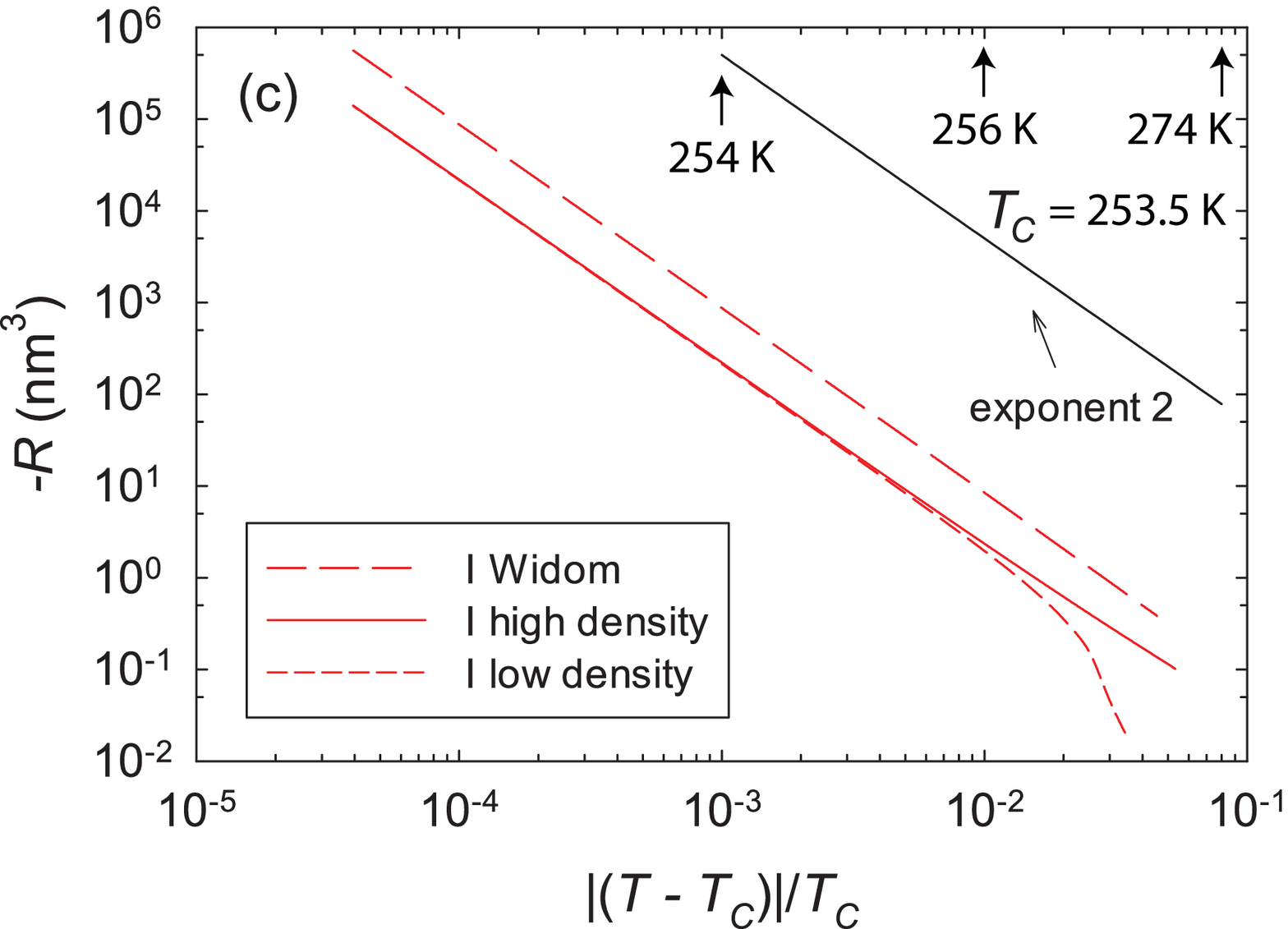}
\end{minipage}
\hspace{0.0 cm}
\begin{minipage}[b]{0.6\linewidth}
\includegraphics[width=2.8in]{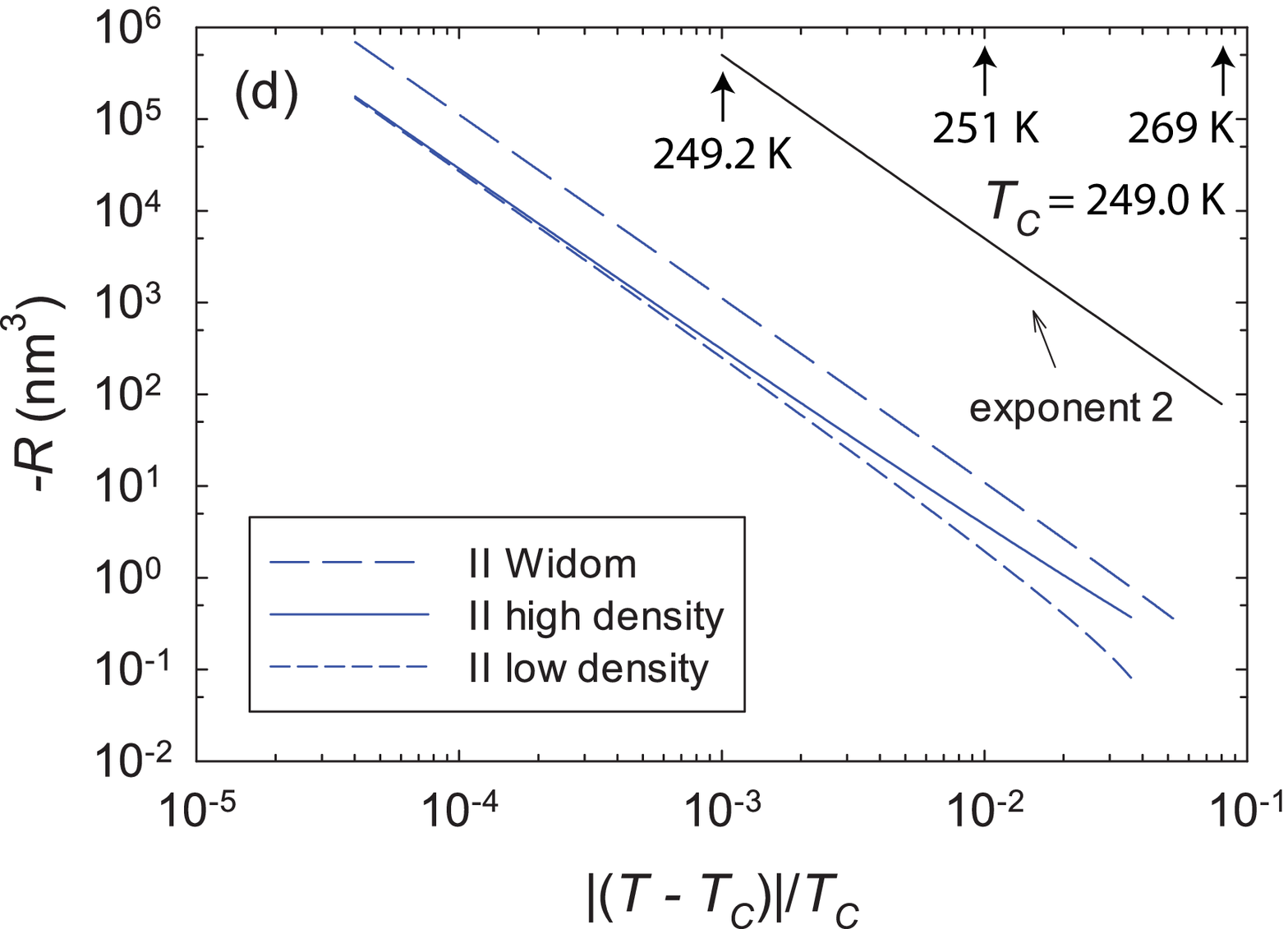}
\end{minipage}
\caption{Log-log plots verifying the asymptotic power law behaviors with MF critical exponents for: (a) $k_{T}$ for ST2(I-MF), and (b) $k_{T}$ for ST2(II-MF). The figures show $k_{T}$ along the Widom lines, and along both the high and low density branches of the LLPT's. Asymptotically, the graphs follow straight lines, with the MF critical exponent 1. Figures (c) and (d) are the corresponding graphs for $-R$, also showing the expected MF behavior, with critical exponent $2$. The temperatures listed along the top axes are for along the Widom line. They are listed purely for guidance.}
\label{figure3}
\end{figure}

Figure \ref{figure3} shows log-log graphs of $k_{T}$ and $-R$ for the ST2(I-MF) and ST2(II-MF) models. Figures \ref{figure3}(a) and \ref{figure3}(b) show $k_{T}$ plotted along the Widom lines and the LLPT's in the high and the low density branches. The linearity of the graphs, with $\gamma=\gamma'=1$, is in accord with MF. Although the power law behavior is expected only asymptotically, the extended linearity of the graphs, most notably along the Widom lines, attests to the reach of the critical point MF theory. Notice that along the LLPT's of both the ST2(I-MF) and ST2(II-MF) models, the fluid phase with the lower density has the smaller compressibility. Thus, the lower density fluid is physically ``harder,'' a finding perhaps unexpected. But this may be just another justification for attributing solid-like properties to the LDL.

\par
Figures \ref{figure3}(c) and \ref{figure3}(d) are the corresponding graphs for $-R$. All the branches show power law behaviors, with MF critical exponent $2$. Striking is the equality of the $-R$ values in the high and low density phases, particularly for the ST2(I-MF) model. This equality is an example of the commensurate $R$ theorem that was used to calculate the phase transition curves for the van der Waals and the Lennard-Jones models \cite{Ruppeiner2012, Ruppeiner2012a,May2012}. Fig. \ref{figure3}(c) looks very similar to Figure 1 for hydrogen in \cite{Ruppeiner2012}, and it is remarkable that we get such close equality for properties in fluids with such different densities. The linear regime extends to small volumes, roughly $10^{-1}$ nm$^3$, corresponding to a sphere with radius about 5 Angstroms.

\par
\begin{figure}
\begin{center} 
\includegraphics[width=4in]{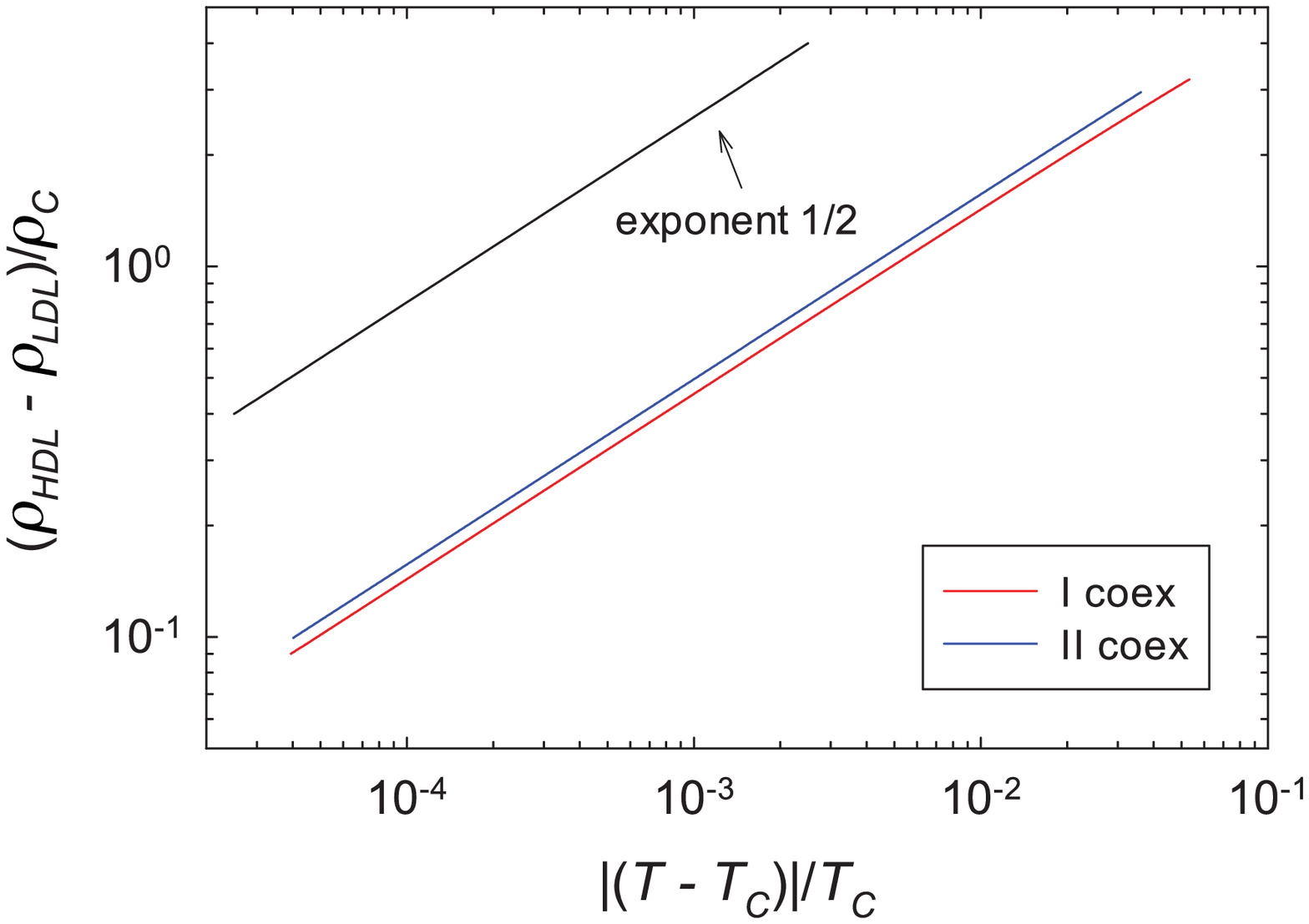} 
\end{center}
\caption{The reduced densities along the LLPT's. The graphs for both the ST2(I-MF) and ST2(II-MF) models are very closely lines with slopes 1/2, in accord with expectations from MF.}
\label{figure4} 
\end{figure}

Figure \ref{figure4} shows the reduced densities $(\rho_{HDL}-\rho_{HDL})/\rho_C$ along the LLPT's. The curves for the two models are closely linear, with exponent $1/2$, in accord with MF.

\subsubsection{The Widom line}

Turn attention again to the Widom line. As we discussed above, the MF picture in play here allows a calculation of the Widom line via the analytic continuation of the LLPT, Eq. (\ref{75}), to $\tau>0$. We would like to compare this analytic continuation method with the alternate method of finding the loci of the maximum correlation lengths. But this later method is usually problematic since the correlation length is not traditionally accessible in thermodynamics. A contribution of the geometry of thermodynamics is that it offers a thermodynamic link to the correlation length via the equality Eq. (\ref{140}), and this is useful in calculating the Widom line \cite{Ruppeiner2012, May2012, Sarkar2013, Brazhkin2014, Corradini2015}.

\par
\begin{figure}
\begin{center} 
\includegraphics[width=4in]{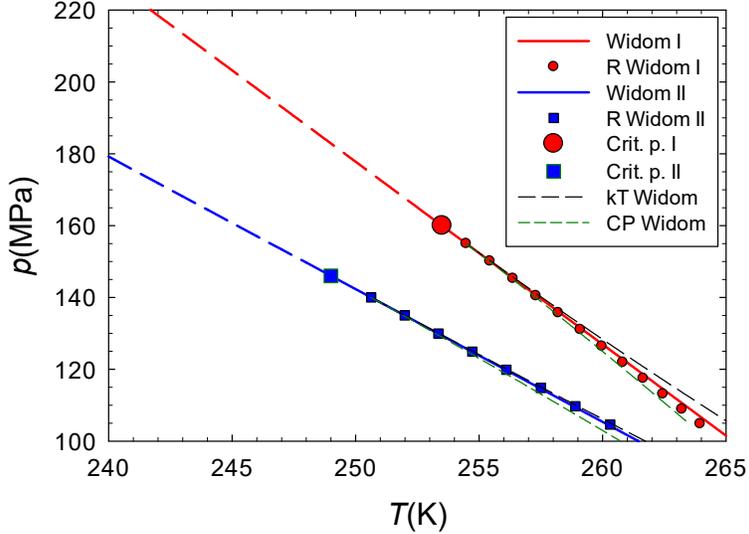} 
\end{center}
\caption{The Widom lines for the ST2(I-MF) and ST2(II-MF) models computed in two ways: (i) by locating the local maxima of $R$ along lines of constant $p$ (red circles: ST2(I-MF), blue squares: ST2(II-MF)), and (ii) by analytic continuation of the LLPT's $\ln K=0$ (solid lines). The methods are in good agreement. The dashed straight lines indicate the LLPT's. For completeness, we also show the Widom lines computed by locating the local maxima of $k_{T}$ and $c_{p}$.}
\label{figure5} 
\end{figure}

We proceeded by finding the locus of curves of maximum $|R|$ along lines of constant pressure, and show results in Figure \ref{figure5}. The two ways of computing the Widom line are seen to be in good agreement with each other. For completeness, we also show the Widom lines computed by locating the local maxima of $k_{T}$ and $c_{p}$.

\section{Conclusion}
Among liquids, water has a number of anomalous properties. For example, the isothermal compressibility $k_{T}$ increases dramatically as liquid water is cooled at constant pressure from the stable phase into the supercooled metastable phase. In contrast, $k_{T}$ for a typical liquid shows no such behavior. The anomalous divergences in water have motivated the picture of a first-order liquid-liquid phase transition line (LLPT), separating two polymorphic phases of liquid water. This LLPT is entirely in the metastable liquid state. It terminates at a critical point (LLCP) expected to be at the heart of the anomalous divergences. In this paper, we focused on an LLPT represented by a two-state mean field (MF) theory equation of state. Data from ST2 simulations based on two different microscopic water models were fit to this MF framework, resulting in the ST2(I-MF) and the ST2(II-MF) models \cite{Holten2014a}.

\par
The polymorphic phases involved in the LLPT are low-density (LDL) and high-density (HDL) liquid states. It is assumed that the LDL state possesses mesoscopic tetrahedral structures that give it solid-like properties, while the HDL is a regular random liquid. But this idea is conjectural since the short-lived nature of these solid-like structures makes them difficult to detect directly. Alternatively, we computed the thermodynamic Ricci curvature scalar $R$, and found significant regimes of positive $R$ in the LDL's of both the ST2(I-MF) and the ST2(II-MF) models, though more pronounced in ST2(I-MF). Positive $R$ is the thermodynamic signature of solid-like properties, so these findings support the proposal of solid-like structures in liquid water. In our paper, we also pointed out a possible linkage between positive $R$ in the supercooled states of ST2(I-MF) and in ambient stable water.

\par
We suggested that the LLPT might fit into the broader context of the phase transition lines of ordinary pure fluids. For pure fluids with molecules not too simple, it was found that there is almost always a regime of positive $R$ in the vapor phase (the low density one). It was suggested that in vapors with densities neither too small nor too large, there is at the same time enough physical space and enough molecular complexity to allow for the organization of groups of molecules into solid-like structures \cite{Ruppeiner2017}. So far, no specific microscopic mechanism has been established as the foundation of this idea, but the similarities of the $R$-diagrams of the LLPT and that of pure fluids is at least suggestive.

\par
In addition, we reported that for the LLPT, $k_{T}$ for the LDL is smaller than that for the coexisting HDL. This is the opposite of what one might naively expect, and is another possible indicator that the ``harder'' LDL has solid-like structures. We found that the density minimum in the LDL for ST2(I-MF) coincides closely with the change in sign of $R$. We also reviewed the theory, verified the MF critical exponents, demonstrated the large reach of the MF critical regime, and calculated the Widom line using $R$.

\par
In conclusion, we have provided here another example showing the nice interpretation that results from looking at $R$.

\subsection*{Acknowledgments}
A large number of ideas in this and other papers were developed in close cooperation with Helge-Otmar May. Since he is gone, his extraordinarily important contributions will be missing in the future.

\pagebreak

\end{document}